**Large language models streamline automated systematic review: A preliminary study**

Xi Chen[1,2] M.D., Ph.D.; Xue Zhang[1,2] P.T.;

1Sports Medicine Center, West China Hospital, West Chian School of Medicine, Sichuan University, Chengdu, Sichuan, China. Email: geteff@wchscu.cn

2Department of Orthopedics and Orthopedic Research Institute, West China Hospital, Sichuan University, Chengdu, Sichuan, China. Email: 2742870675@qq.com

Large Language Models (LLMs) have shown promise in natural language processing tasks, with the potential to automate systematic reviews. This study evaluates the performance of three state-of-the-art LLMs in conducting systematic review tasks. We assessed GPT-4, Claude-3, and Mistral 8x7B across four systematic review tasks: study design formulation, search strategy development, literature screening, and data extraction. Sourced from a previously published systematic review, we provided reference standard including standard PICO (Population, Intervention, Comparison, Outcome) design, standard eligibility criteria, and data from 20 reference literature. Three investigators evaluated the quality of study design and eligibility criteria using 5-point Liker Scale in terms of accuracy, integrity, relevance, consistency and overall performance. For other tasks, the output is defined as accurate if it is the same as the reference standard. Search strategy performance was evaluated through accuracy and retrieval efficacy. Screening accuracy was assessed for both abstracts screening and full texts screening. Data extraction accuracy was evaluated across 1,120 data points comprising 3,360 individual fields. Claude-3 demonstrated superior overall performance in PICO design (overall performance: GPT-4: 4.25±0.45 vs Claude-3: 4.58± 0.52, P=0.039; GPT-4: 4.25±0.45 vs Mistral: 3.58±0.67, P=0.016; Claude-3 vs Mistral, P=0.002). In search strategy formulation, GPT-4 and Claude-3 achieved comparable accuracy, outperforming Mistral (GPT-4: 0.78 vs Claude-3: 0.79 vs Mistral: 0.67, P=0.096). For abstract screening, GPT-4 achieved the highest accuracy, followed by Mistral and Claude-3 (GPT-4: 0.92 vs Mistral: 0.84, P<0.001; GPT-4: 0.92 vs Claude-3: 0.69, P<0.001; Mistral vs Claude-3, P<0.001). In data extraction, GPT-4 significantly outperformed other models (accuracy: GPT-4: 0.81 vs Claude-3: 0.59, P<0.001; GPT-4: 0.81 vs Mistral: 0.46, P<0.001, Claude-3 vs Mistral, P<0.001). LLMs demonstrate potential for automating systematic review tasks, with GPT-4 showing superior performance in search strategy formulation, literature screening and data extraction. These capabilities make them promising assistive tools for researchers and warrant further development and validation in this field.

**Introduction**

Large Language Models (LLMs) are a type of deep learning-based natural language processing model that can generate human-like text, and perform natural language processing tasks[1]. LLMs have shown promising performance in a wide range of applications particularly in medicine[2-4].

Systematic reviews synthesize relevant evidence on a specific research question or topic[5], and provide best guidance for clinical decision-making[6]. A systematic review workflow mainly consists of study design, literature search, screening and data extraction, all of which require diverse expertise and access to databases and tools[7]. The process often takes over a year[8], making it time-consuming, labor-intensive and prone to human-error.

Automation in systematic reviews has gained interest, with researchers exploring various tools and technologies to streamline the review process[9]. O'Mara-Eves and Marshall et al explored ways to streamline and automate various stages of the systematic review process to reduce the workload on researchers, minimize human error, and accelerate the production of high-quality systematic reviews[10,11]. However, the machine learning approaches face limitation in data extraction and handing complex tasks[12].

Beyond that, LLMs have shown promise in systematic review with Lai et al. finding that ChatGPT and Claude perform well in ROB assessment[13]. Feng and Thomas et al. identified that text mining techniques are useful for automated document classification and summarization, aiding in search term identification, study selection, and data extraction[14,15].

There were still some defects for LLMs to automate a SR. LLMs show potential in abstract screening but their prospective performance in full-text screening for systematic reviews remains unclear[16-18]. While tools like ChatGPT can assist in parts of the review process, errors persist[19]. In addition, no research compared the performance of different LLMs in conducting systematic reviews.

Advances in models like GPT-4 and Claude-3, along with techniques like prompt engineering and retrieval-augmented generation, have improved LLM capabilities.[20,21,22].

Integrating these advancements offers potential for automating the entire systematic review process.

This study aims to evaluate and compare the performance of GPT-4, Claude-3, and Mistral 8x7B in streamlining systematic review tasks, including study design, search strategies, screening, and data extraction, to assess their potential and limitations.

**Methods**

Study design

This study aims to use prompt engineering to enable large language models (LLMs) to perform tasks in systematic review including PICO (Population, Intervention, Comparison, Outcome) design and eligibility criteria for given research questions, formulating search strategies, screening literature, and extracting data。 We evaluate the performance of three state-of-the-art large language models (LLMs)—GPT-4, Claude-3, and Mistral 8x7B—in conducting systematic reviews.

Dataset

A previously published systematic review and the clinical trials included in it were curated into a test dataset. Raw data from the systematic review, as well as its registration protocol, were retrieved, including the original research question, PICO framework, eligibility criteria, search terms, search results, abstract screening results, full-text screening results, and extracted data. The clinical trials included in the systematic review were retrieved and converted into JSON format, with figures removed using the PyMuPDF package in Python. The dataset was examined by two researchers for accuracy and was curated into a test set and a ground-truth set for subsequent testing and evaluation.

Large Language Models Testing

models, API, deployment parameters

Three large language models—GPT-4, Claude-3, and Mistral 8x7B—were selected due to their state-of-the-art performance across various evaluation metrics[23–25]. The specific model versions tested in this study were Claude 3 (anthropic.claude-3-sonnet-20240229-v1:0), GPT-4 (gpt-4-turbo-preview), and Mistral 8x7B (mistral.mixtral-8x7b-instruct-v0:1). All models were accessed through Application Programming Interfaces (APIs) in independent sessions. For all tested models, the temperature parameter was set to zero to ensure more reliable and reproducible results[26]. The language models' outputs were recorded as JSON files.

Prompt Development

Distinct testing prompts were developed for each task. All prompts were iteratively developed and modified until they consistently adhered to the task instructions and produced output in the required format. For literature screening and data extraction, the prompts were developed during a pilot test using two randomized controlled trials.

LLMs' automation tasks and evaluation framework

LLMs automated the following tasks of SR: formulate study design based on a specific research question, formulate search strategies, screen literature, and extract data. The research question used for SR automation in this study is *How does neuraxial anaesthesia compare to general anaesthesia in terms of adverse events risk in patients undergoing hip fracture surgery*, which is based on a previous systematic review we published. The information from this prior SR serves as the foundational data for LLMs to perform the relevant tasks and as the reference standard during evaluation. The LLMs automation approach is illustrated in **Figure 1**. The detailed description of each task is as follows:

**Figure 1.** Design of the study.

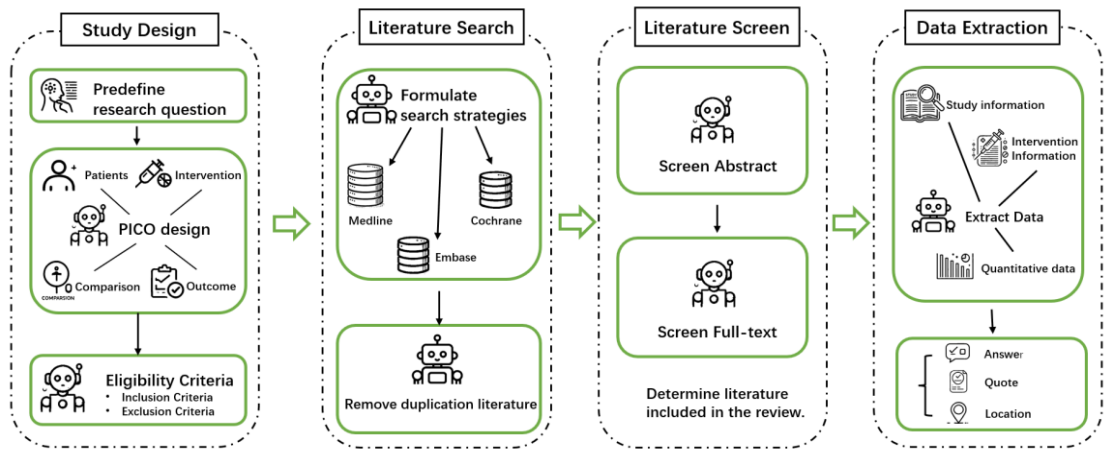

**Study Design:** When a research question (such as general anaesthesia superior to neuraxial anaesthesia in patients undergoing hip fracture surgery) was provided, LLMs would fomulate the PICO of the study and develop eligibility criteria for literature screening.

One independent investigator involved in the previous systematic review assessed the quality of PICO design and eligibility criteria based on reference standard three times. They evaluated model performance based on four metrics: accuracy, integrity, relevance and consistency. Each metric was measured based on a 5-point Liker Scale[27]. The researcher evaluated the answers blinded to the LLM that generated the answer and reached a consensus score through discussion. The detailed quality assessment criteria are listed in **Supplementary File 1**.

**Search Strategies:** Large Language Models (LLMs) were provided with standard PICO and eligibility criteria and were tasked with formulating search terms for MEDLINE, Embase, and the Cochrane Library using this information.

The evaluation was performed by comparing the search results yielded by the search terms generate by LLMs with the search results from the reference standard. First, we evaluated whether the search terms adhered to the corresponding database's requirements and successfully yielded search results. We then evaluated whether the search terms accurately retrieved the relevant publications that should be included in the final SR. We also assessed the total number and accuracy of retrieved publications to evaluate the efficiency of the search strategies.

**Literature screen:** Large Language Models (LLMs) were tasked with screening both abstracts and full-text articles. In the reference systematic review, 811 publications underwent title and abstract screening, of which 29 publications proceeded to full-text screening. Using identical eligibility criteria in the published SR, LLMs performed screening on the same set of publications.

The screening performance was evaluated by comparing LLMs' decisions with the reference standard for each publication. A screening decision was considered accurate if it matched the reference standard's inclusion or exclusion determination. Accuracy was calculated as the ratio of correct screening decisions to the total number of publications screened, and was assessed separately for both abstract and full-text screening phases.

**Data collection:** LLMs were tasked to extract data from full-text articles, which included study information, intervention information, and outcome information. Study information included: first author, publishing year, country, planned sample size, final sample size, primary outcome, secondary outcomes, and follow-up period. Intervention information included: neuraxial anaesthesia technique, sedation technique used in neuraxial anaesthesia, induction technique in general anaesthesia, airway management technique in general anaesthesia, and maintenance technique in general anaesthesia. Outcome information included the following outcomes: mortality, delirium, heart failure, hospital stay, pulmonary embolism, postoperative nausea and vomiting, surgery time, pneumonia, acute kidney injury, and cerebral vascular accident. For each outcome, the following parameters were extracted: presence of outcome, outcome measurement time point, outcome value in general anaesthesia group, and outcome value in neuraxial anaesthesia group. For each data point, three elements were extracted: answer, quote, and location. Answer referred to the response of the LLMs responding to the extraction task. Quote referred to the original content in the articles which contained the required information. Location referred to the section and paragraph of the quote within the original article. For example, in extracting the postoperative delirium rate for the general anaesthesia group from one article, the output was structured as follows: Answer: 6/55 (5.5%); Location: Table 2: Postoperative outcomes section; Quote: Postoperative delirium: 6/55 (5.5%).

Data verification was conducted independently by two researchers. Given that correct answers for a single data point could be extracted from various sections of the original text, the evaluation process involved comparing the model outputs directly with the source documents. For each model, a total of 1,120 data points, comprising 3,360 individual data fields, were evaluated. A total of 10,080 data fields were manually examined. Finally, the accuracy was calculated by the ratio of the number of correctly extracted data points to the total number of data points that should be extracted.

**Reliability analysis**

In this study, we evaluated the reliability of three LLMs in conducting systematic reviews across the main tasks: study design, search strategy formulation, and literature screening. Each model performed each task five times to assess the consistency of its performance. The Intraclass Correlation Coefficient (ICC) and Coefficient of Variation (CV) were calculated to quantify the models' reliability across repeated runs[28,29].

**Statistical Analysis**

All statistical analyses were performed using the SPSS 25.0 software (IBM, Armonk, NY, USA) and GraphPad Prism version 8 (GraphPad Software, San Diego, CA, USA). Continuous variables were calculated by means and standard deviations and analysed using Kruskal-Wallis H test. Discontinuous data were expressed as incidence and rate and analysed using the chi-square test for differences. A P value less than 0.05 indicated statistical significance. A post-hoc analysis was conducted using pairwise comparisons with significance levels adjusted by Bonferroni correction. After adjustment, the threshold for statistical significance was set at 0.0167. Therefore, for the post-hoc analysis, a p-value of less than 0.0167 indicates a statistically significant difference between groups.

**Results**

**Evaluation of LLMs' performance**

**Study design**

For PICO design, in accuracy, the score of GPT-4, Claude-3 and Mistral8x7B were 4.33 ±0.58, 4.67±0.58 and 3.33±0.58 (P=0.100). In integrity, the score of GPT-4, Claude-3 and Mistral8x7B were 4, 4.33±0.58 and 3.67±0.58 (P=0.264). In relevance, the score of GPT-4, Claude-3 and Mistral8x7B were 4.33±0.58, 4.67±0.58 and 4.33±0.58 (P=0.670). In consistency, the score of GPT-4, Claude-3 and Mistral8x7B were 4, 4.33±0.58 and 3.33±0.58 (P=0.110). the overall performance score of Claude-3 was significantly higher than GPT-4 and Mistral8x7B (GPT-4: 4.25 ±0.45 vs Claude-3: 4.58±0.52, P=0.039; GPT-4: 4.25±0.45 vs Mistral: 3.58±0.67, P=0.016; Claude-3 vs Mistral, P=0.002). **Table 1** and **Figure 2a-e** show the evaluation results of four metrics for PICO design. **Figure 2f** shows the PICO design examples generated by the three LLMs. **Figure 2g** shows the examples of investigators assessing the PICO design.

**Table 1.** Evaluation of PICO Design

|  | GPT-4 | Claude-3 | Mistral8x7B | P Value |
| --- | --- | --- | --- | --- |
| Accuracy | 4.33 (0.58) | 4.67 (0.58) | 3.33 (0.58) | 0.100 |
| Integrity | 4 | 4.33 (0.58) | 3.67(0.58) | 0.264 |
| Relevance | 4.33 (0.58) | 4.67 (0.58) | 4.33 (0.58) | 0.670 |
| Consistency | 4 | 4.33(0.58) | 3.33 (0.58) | 0.110 |
| Overall Performance | 4.25 (0.45) | 4.58 (0.52) | 3.58 (0.67) | 0.002[#] |

Scores are presented as mean (SD). # GPT-4 vs Claude-3 P=0.039; GPT-4 vs Mistral P=0.016; GPT-4 vs Mistral P=0.002

**Figure 2.** Performance of LLMs generating PICO design.

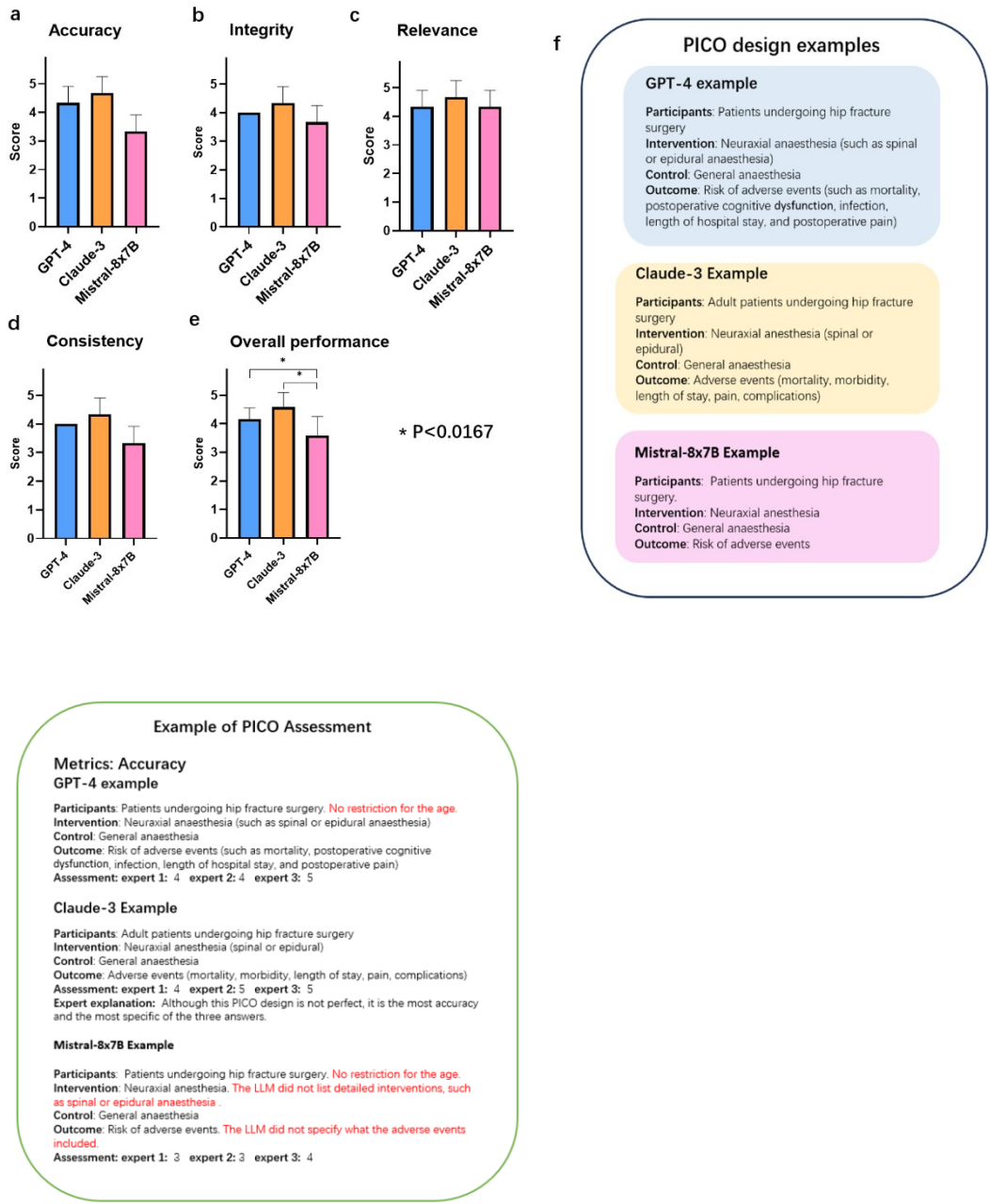

a. scores on accuracy of the PICO design; b. scores on integrity of the PICO design; c. scores on relevance of the PICO design; d. scores on consistency of the PICO design; e. the overall performance of the PICO design; f. examples of PICO design generated by GPT4, Claude-3 and Mistral7×8B; g. examples of PICO design assessment.

For Eligibility criteria, the score of GPT-4, Claude-3 and Mistral8x7B were 3.67±0.58, 3.67±0.58 and 4 (P=0.11) respectively in accuracy. The score of GPT-4, Claude-3 and Mistral8x7B were 3.67±0.58, 4.33±0.58 and 4±1.0 (P=0.513) respectively in integrity. The score of GPT-4, Claude-3 and Mistral8x7B were 3.67±0.58, 4.33±0.58 and 4.33±0.58 (P=0.304) in relevance. In consistency, the score of GPT-4, Claude-3 and Mistral8x7B were 4.33±0.58, 3.67±0.58 and 4.33±0.58 (P=0.304). The overall performance score of GPT-4, Claude-3 and Mistral8x7b were 3.75±0.62, 4.33±0.65, 4.25±0.45 respectively. The Claude-3 outperformed in the three LLMs (p=0.05). Table 2 and **Figure 3a-e** demonstrate the evaluation results for eligibility criteria. **Figure 3f** shows the PICO design examples generated by the three LLMs. Examples of investigators assessing the eligibility criteria was summarized in the **Figure 3g**.

**Table 2.** Evaluation of Eligibility Criteria

|  | GPT-4 | Claude-3 | Mistral8x7b | P Value |
|---|---|---|---|---|
| Accuracy | 3.67 (0.58) | 4.67 (0.58) | 4 | 0.110 |
| Integrity | 3.67 (0.58) | 4.33 (0.58) | 4 (1.0) | 0.513 |
| Relevance | 3.67 (0.58) | 4.33(0.58) | 4.33 (0.58) | 0.304 |
| Consistency | 4.33 (0.58) | 3.67 (0.58) | 4.33 (0.58) | 0.304 |
| Overall Performance | 3.75 (0.62) | 4.33(0.65) | 4.25(0.45) | 0.05 |

Scores are presented as mean (SD).

**Figure 3**. Performance of LLMs generating eligibility criteria.

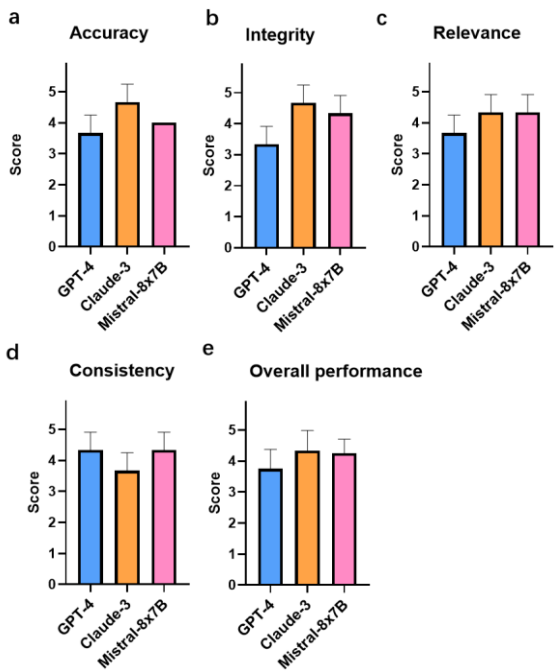

a. accuracy of the eligibility criteria; b. integrity of the eligibility criteria; c. relevance of the eligibility criteria; d. consistency of the eligibility criteria; e. overall performance of the eligibility criteria; f. examples of eligibility criteria generated by GPT4, Claude-3 and Mistral7×8B; g. examples of eligibility criteria assessment.

**Search strategies.**

Representative search terms generated by GPT-4, Claude-3, and Mistral-7B×8 for MEDLINE, Embase, and Cochrane databases are presented in **Figure 4a**. An error in the MEDLINE search terms generated by Mistral-7B×8 is illustrated in **Figure 4b**. The overall accuracy of search terms generated by GPT-4, Claude-3 and Mistral were 0.78, 0,79, 0.67 (P=0.096), respectively. The overall number of publications retrieved by GPT-4, Claude-3 and Mistral were 719, 591, 518, respectively. The respective accuracy of search terms and the overall number of publications retrieved by GPT-4, Claude-3 and Mistral are shown in **Table 3**.

**Figure 4**. Performance of search strategies formulation.

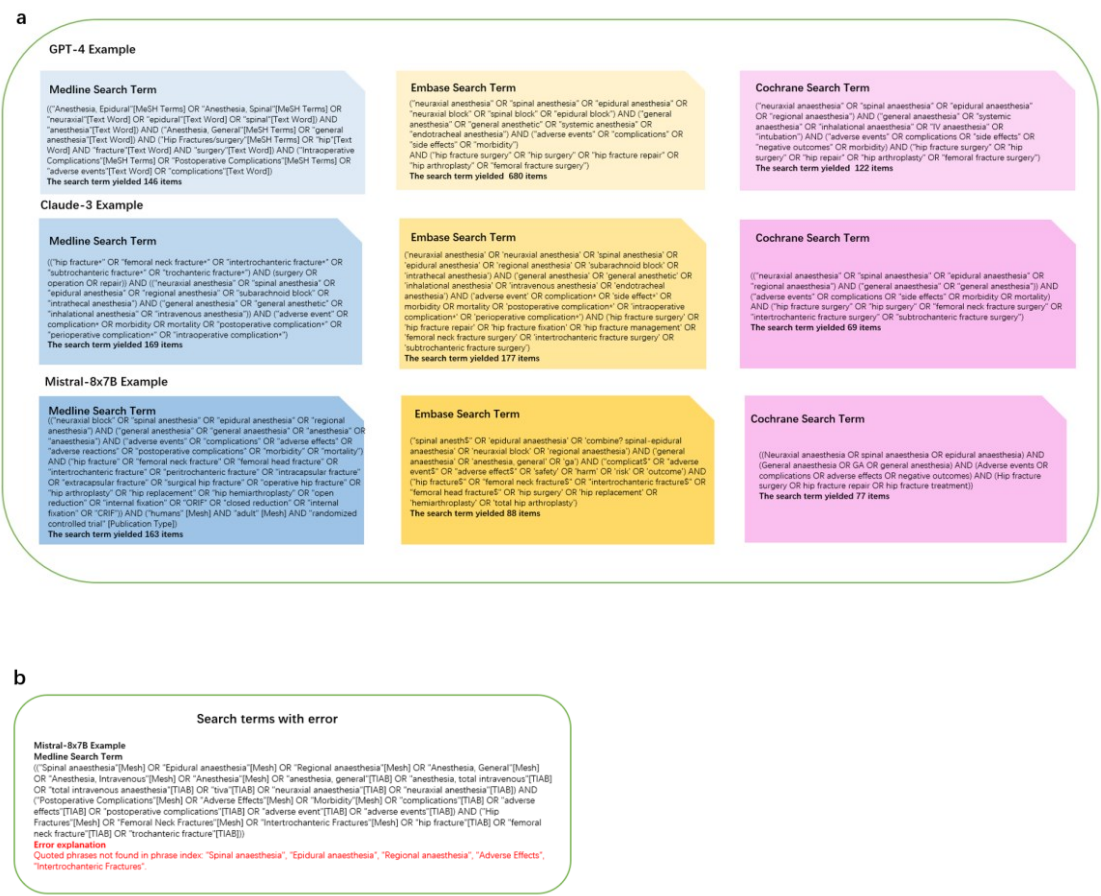

a. examples of search terms generated by GPT-4, Claude-3 and Mistral7×8B for Medline, Embase and Cochrane.; b. example of search terms with error; c. accuracy and search results of the search strategies for 5-times retrieval.

**Table 3.** Accuracy and Number of Publications Retrieved for the Search Strategies

|  | GPT-4 | Claude-3 | Mistral8x7B | P Value |
|---|---|---|---|---|
| Accuracy | 0.78 | 0.79 | 0.67 | 0.096 |
| Number of Publications Retrieved | 719 | 591 | 518 | - |

**Literature screening**

The abstract screening accuracy of GPT-4, Claude-3 and Mistral7×8B were 0.92, 0.69, 0.84 and showed significant differences (GPT-4 vs Mistral, P<0.001; GPT-4 vs Claude-3, P<0.001; Mistral vs Claude-3, P<0.001). **Figure 5a** demonstrate the accuracy of correct screening results for abstract screening, respectively. The example of incorrect abstract screening is demonstrated in **Figure 5c**. The full-text screening accuracy for each model was 0.75, 0.69 and 0.79, which is demonstrated in **Figure 5b**. Examples of incorrect full-text screens are summarized in **Figure 5d**.

**Figure 5**. Performance of literature screening.

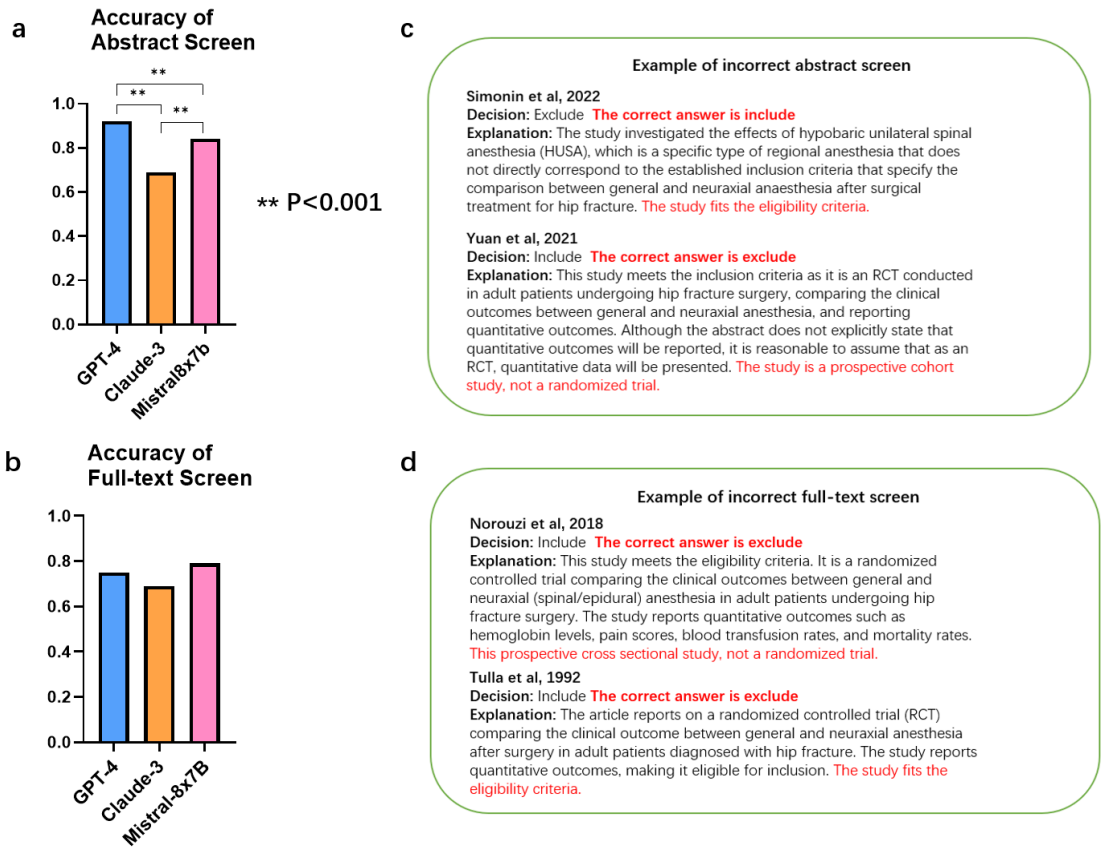

a. accuracy of abstract screen; b. accuracy of full-text screen. c. example of incorrect abstract screen; d. example of incorrect full-text screen;

**Data extraction**

For each data point, three key elements were examined: the answer, the quote, and the location within the text. For each model, a total of 1,120 data points, comprising 3,360 individual data fields, were evaluated. For the overall accuracy, GPT-4 significantly outperformed among the three LLMs (GPT-4, 0.81 vs Claude-3, 0.59 vs Mistral7×8B, 0.46, P<0.001). Additionally, GPT-4 also had a more prominent performance than Claude-3 and Mistral7×8B in terms of three key elements: answer (GPT-4, 0.81 vs Claude-3, 0.60 vs Mistral7×8B, 0.47, P<0.001), quote (GPT-4, 0.80 vs Claude-3, 0.54 vs Mistral7×8B, 0.41, P<0.001) and location (GPT-4, 0.83 vs Claude-3, 0.62 vs Mistral7×8B, 0.49, P<0.001). The results are showed in **Table 3** and **Figure 6a-d**.

**Table 4.** Overall accuracy of the Data extraction

|          | GPT-4 | Claude-3 | Mistral8x7B | P Value  |
|----------|-------|----------|-------------|----------|
| Answer   | 0.81  | 0.60     | 0.47        | <0.001[a] |
| Quote    | 0.80  | 0.54     | 0.41        | <0.001[a] |
| Location | 0.83  | 0.62     | 0.49        | <0.001[a] |
| Total    | 0.81  | 0.59     | 0.46        | <0.001[a] |

[a] Further analysis showed all pairwise comparison had P value less than 0.001;

**Figure 6**. Overall accuracy of the data extraction.

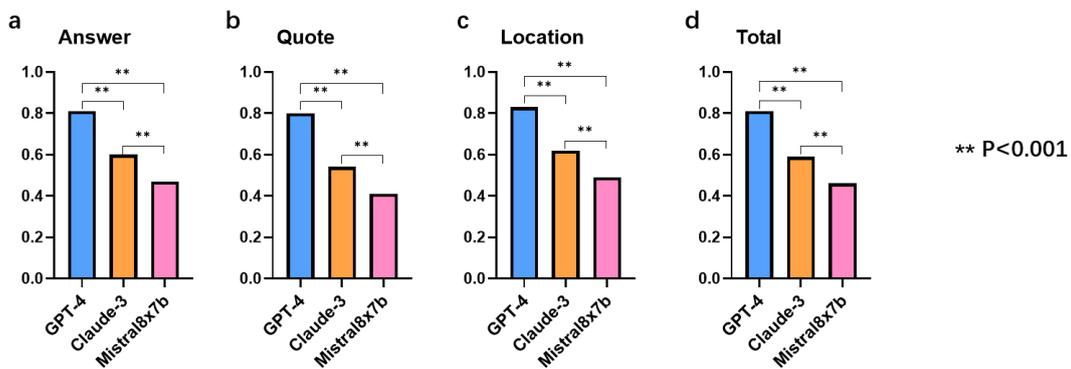

a. Answer; b. Quote; c. Location. d. Overall;

In study information extraction, the total accuracy of GPT-4, Claude-3 and Mistral7×8B were 0.85, 0.80, 0.64, (P<0.001). Meanwhile, GPT-4 achieved higher accuracy than Claude-3 and Mistral7×8B for the answer (0.81 vs 0.77 vs 0.62, P=0.376), quote (0.84 vs 0.78 vs 0.66, P<0.05), and location (0.91 vs 0.86 vs 0.63, P<0.001). The results are showed in **Table 4** and **Figure 7**.

**Table 5.** Accuracy of Study Information extraction.

|          | GPT-4 | Claude-3 | Mistral8x7B | P Value    |
|----------|-------|----------|-------------|------------|
| Answer   | 0.81  | 0.77     | 0.62        | 0.376[e]   |
| Quote    | 0.84  | 0.78     | 0.66        | 0.003[f]   |
| Location | 0.91  | 0.86     | 0.63        | <0.001[d]  |
| Total    | 0.85  | 0.80     | 0.64        | <0.001[d]  |

**Figure 7.** Performance of LLMs extracting study information.

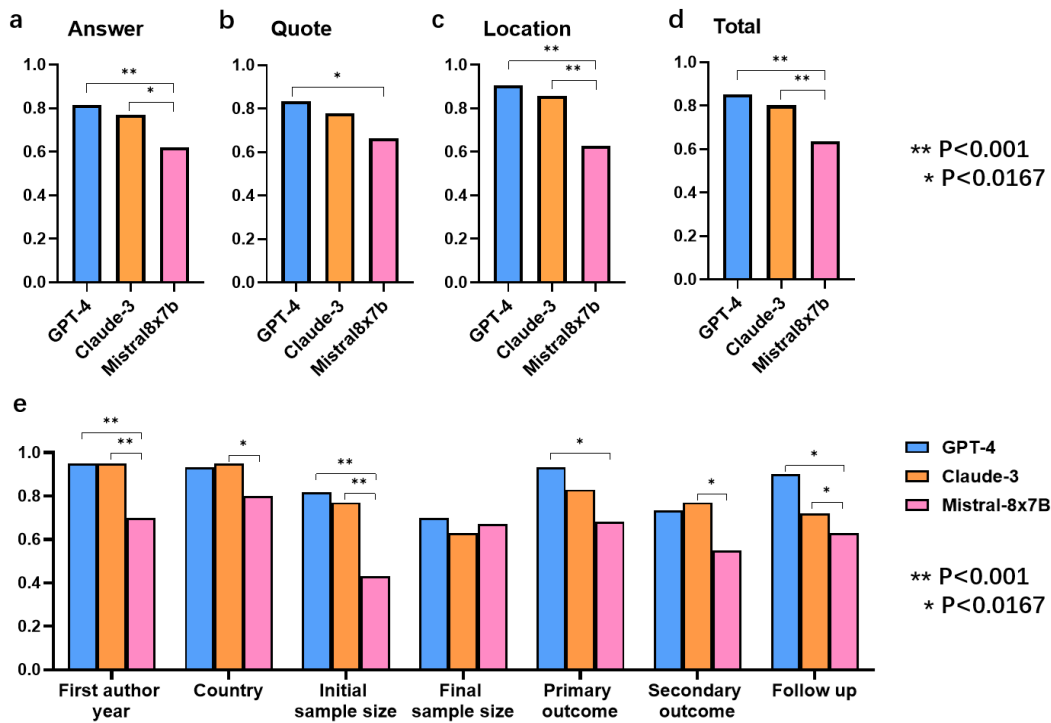

a. accuracy of extracting answer in study information; b. accuracy of extracting quote in study information; c. accuracy of extracting location in study information. d. overall accuracy of extracting study information; e. accuracy of each item extraction within study information category.

In intervention information extraction, the overall accuracy of GPT-4, Claude-3 and Mistral7×8B were 0.92, 0.94, 0.82, (P<0.001). Claude-3 and GPT-4 outperformed Mistral in the answer (GPT-4, 0.90 vs Claude-3, 0.90 vs Mistral7×8B, 0.75, P<0.05), quote (GPT-4, 0.92 vs Claude-3, 0.95 vs Mistral7×8B, 0.83, P<0.05), and location (GPT-4, 0.94 vs Claude-3, 0.98 vs Mistral7×8B, 0.87, P<0.05). The results are showed in **Table 5** and **Figure 8**.

**Table 6.** Accuracy of Intervention Information Extraction.

|  | GPT-4 | Claude-3 | Mistral8x7B | P Value |
| --- | --- | --- | --- | --- |
| Answer | 0.90 | 0.90 | 0.75 | 0.003 |
| Quote | 0.92 | 0.95 | 0.83 | 0.013 |
| Location | 0.94 | 0.98 | 0.87 | 0.009 |
| Total | 0.92 | 0.94 | 0.82 | <0.001[d] |

[d] Further analysis showed P<0.001 for GPT-4 vs Claude-3, P<0.001 for GPT-4 vs Mistral8x7B.

**Figure 8.** Performance of LLMs extracting intervention information.

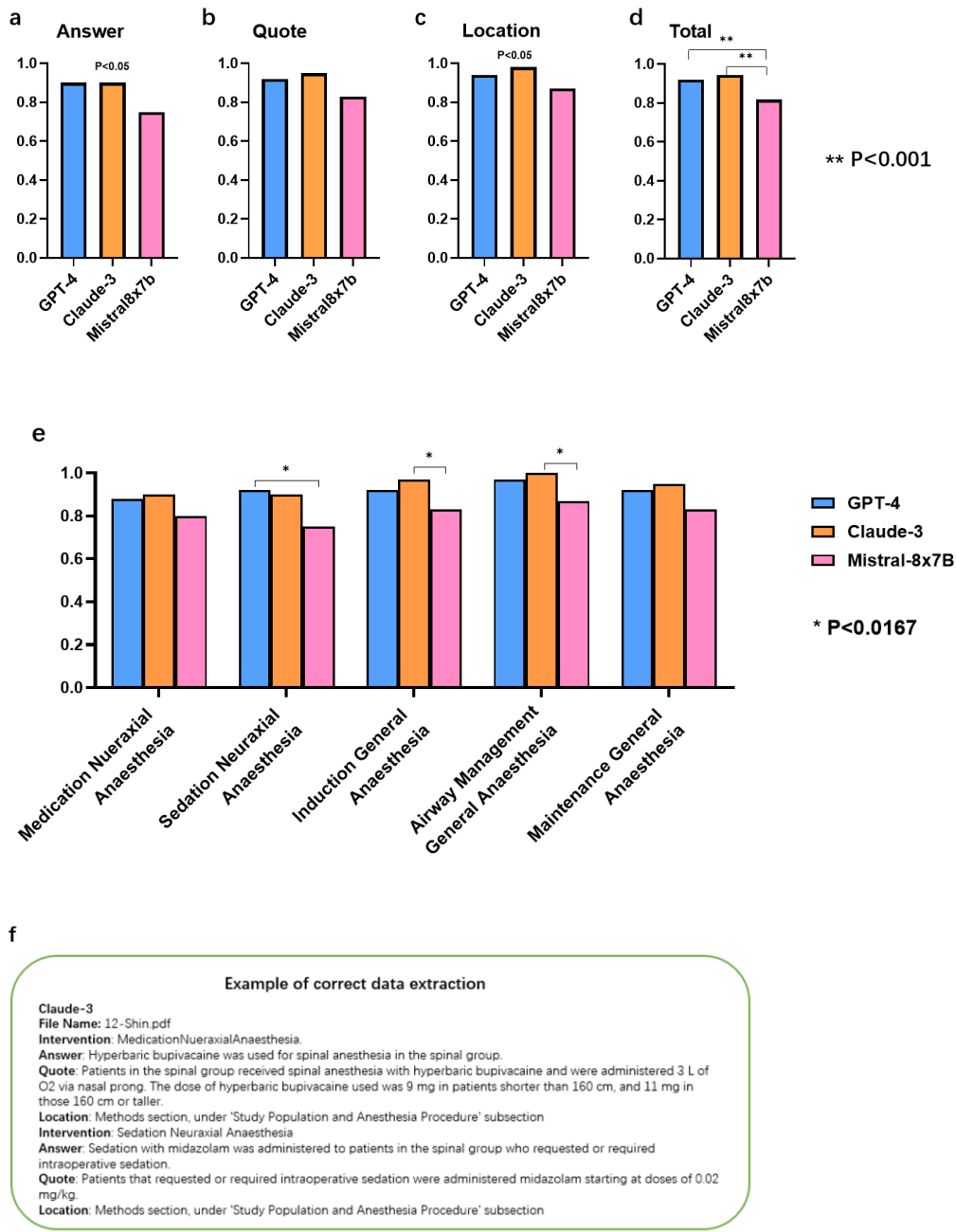

a. accuracy of extracting answer in intervention information; b. accuracy of extracting quote in intervention information; c. accuracy of extracting location in intervention information. d. overall accuracy of extracting intervention information; e. accuracy of each item extraction within intervention information category. f. example of correct data extraction.

In quantitative outcome extraction, the accuracy of GPT-4, Claude-3 and Mistral7×8B were 0.80, 0.51, 0.39, (P<0.001). GPT-4 significantly outperformed other models in answer (GPT-4, 0.80 vs Claude-3, 0.54 vs Mistral7×8B, 0.42, P<0.001), quote (GPT-4, 0.78 vs Claude-3, 0.46 vs Mistral7×8B, 0.33, P<0.001) and location (GPT-4, 0.80 vs Claude-3, 0.54 vs Mistral7×8B, 0.42, P<0.001). The results are showed in **Table 6** and Figure **9**. The extraction accuracy for each quantitative outcome are showed in **Table 7** and **Figure 9e**.

**Table 7.** Accuracy of outcome information extraction.

|  | GPT-4 | Claude-3 | Mistral8x7B | P Value |
| --- | --- | --- | --- | --- |
| Answer | 0.80 | 0.53 | 0.42 | <0.001[a] |
| Quote | 0.78 | 0.46 | 0.33 | <0.001[a] |
| Location | 0.80 | 0.54 | 0.42 | <0.001[a] |
| Total | 0.80 | 0.51 | 0.39 | <0.001[a] |

[a] Further analysis showed all pairwise comparison had P value less than 0.001;

**Figure 9.** Performance of LLMs extracting outcome information.

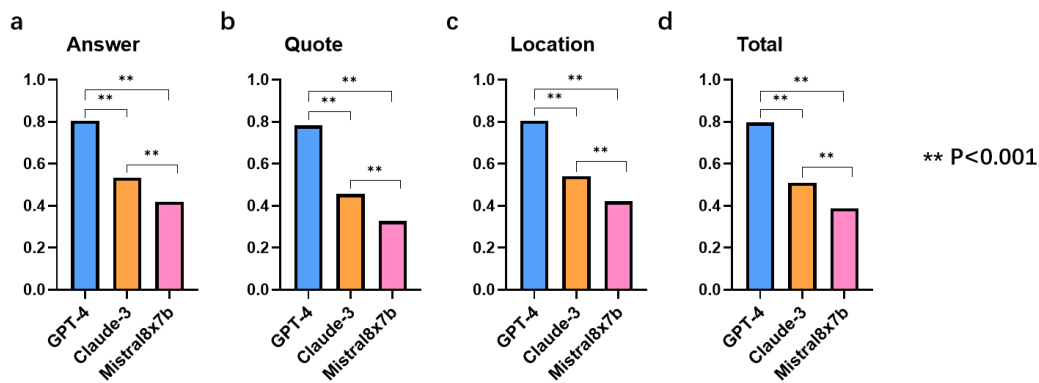

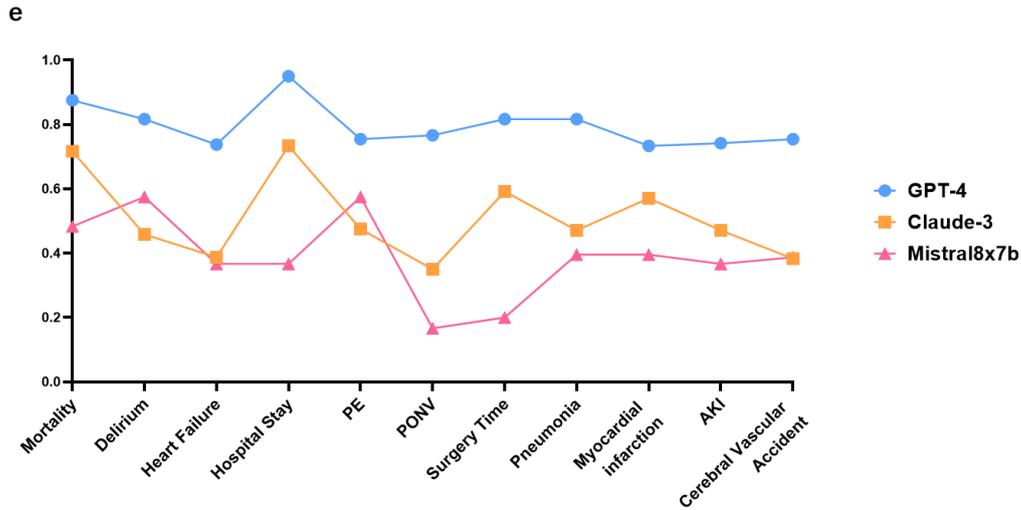

a. accuracy of extracting answer in outcome information; b. accuracy of extracting quote in outcome information; c. accuracy of extracting location in outcome information. d. accuracy of each item extraction within intervention information category. f. example of incorrect data extraction.

Table 8. Accuracy of specific outcome extraction

|  | GPT-4 | Claude-3 | Mistral8x7B | P Value |
| --- | --- | --- | --- | --- |
| Cerebral Vascular Accident | 0.89 | 0.72 | 0.48 | <0.001[a] |
| Delirium | 0.82 | 0.46 | 0.58 | <0.001[b] |
| Heart Failure | 0.74 | 0.39 | 037 | <0.001[c] |
| Hospital Stay | 0.95 | 0.73 | 037 | <0.001[a] |
| PE | 0.75 | 0.48 | 0.58 | <0.001[c] |
| PONV | 0.77 | 0.35 | 0.17 | <0.001[a] |
| Surgery Time | 0.82 | 0.59 | 0.20 | <0.001[a] |
| Pneumonia | 0.82 | 0.47 | 0.40 | <0.001[c] |
| Myocardial infarction | 0.73 | 0.57 | 0.40 | <0.001[a] |
| AKI | 0.74 | 0.47 | 0.37 | <0.001[c] |
| Cerebral Vascular Accident | 0.75 | 0.38 | 0.39 | <0.001[c] |

[a] Further analysis showed all pairwise comparison had P value less than 0.001;

[b] Further analysis showed P value<0.0167 for Claude-3 vs Mistral8x7B, <0.001 for GPT-4 vs Claude-3, <0.001 for GPT-4 vs Mistral8x7B.

[c] Further analysis showed P value>0.0167 for Claude-3 vs Mistral8x7B, <0.001 for GPT-4 vs Claude-3, <0.001 for GPT-4 vs Mistral8x7B.

**Reliability Analysis**

The reliability of study design evaluation was evaluated by Intraclass Correlation Coefficient (ICC). The ICC analysis showed good agreement (≥0.75) in 4 out of 8 evaluation, moderate agreement (≥0.5) in 3 out of 8 evaluation and poor agreement (0.375) in the remaining one. The coefficients of variation (CVs) were calculated for the accuracy of GPT-4, Claude-3, and Mistral 7×8B across five repetitions for each task. GPT-4 and Claude-3 showed low variability (CV<0.10) search strategies and literature screen, which indicated the LLMs achieved consistent and stable performance in executing tasks. Mistral 7×8b demonstrated low variability (CV<0.10) in literature screen but moderate variability (CV=0.136) in search strategies. Detailed results are listed in **Supplementary File 2**.

**Discussion**

This study evaluated the performance of three LLMs (GPT-4, Claude-3, and Mistral8x7B) in conducting systematic reviews across multiple tasks. Claude-3 demonstrated superior performance in PICO design and eligibility criteria evaluation. In formulating search strategies, GPT-4 and Claude-3 showed similar accuracy and both outperformed Mistral8x7B.

For literature screening, GPT-4 achieved the highest accuracy in abstract screening. In data extraction, GPT-4 significantly outperformed the other models in overall accuracy, particularly excelling in quantitative outcome extraction. Both GPT-4 and Claude-3 demonstrated consistent performance with low variability across tasks, while Mistral8x7B showed moderate variability in search strategies. These findings suggest that while different LLMs may excel in specific aspects of systematic review tasks, GPT-4 and Claude-3 generally demonstrate more reliable and superior performance compared to Mistral8x7B. The results indicate that LLMs, particularly GPT-4 and Claude-3, could potentially serve as valuable assistive tools in systematic review processes, though human oversight remains essential for ensuring accuracy and quality.

Currently, no studies have explored or evaluated the performance of large language models in study design. However, there has been considerable exploration into the understanding and generation capabilities of large language models within specialized knowledge domains. Some researches found that the Large Language Model chatbot appeared capable of understanding and responding to long user-written eye health posts, radiation oncology questions and laboratory test questions, and then largely generated appropriate responses that did not differ significantly from ophthalmologist-written responses or experts answers in terms of incorrect information, likelihood of harm, extent of harm, or deviation[30-32]. In our study, regarding the PICO design, all three models demonstrated a strong understanding of the research question and generated satisfactory results. However, in terms of generating eligibility criteria, large language models exhibited limitations in integrating the research question with practical application. As illustrated by **Figure 3g**, LLMs failed to restrict patients' age and specify the types of studies, nor did they require that included studies report at least one quantitative outcome. If applied directly, these eligibility criteria may lead to an excessive workload during subsequent literature screening.

Structuring search strategies in different databases by different Large Language Models (LLMs) has not been critically evaluated yet[33]. But when Nathalia et al. used ChatGPT to generate search strategies for the MEDLINE database, they found AI does not insert the synonymous terms (Entry Terms) and the jargon used in the clinical practice of the researchers, nor organize, in a correct manner, the groups of acronyms in the same search key[33]. Wang and collaborators perceived that

ChatGPT had the potential to generate effective Boolean queries and lead to high search precision, although ChatGPT generates different queries even if the same prompt is used, which vary in effectiveness[34]. Existing researches have acknowledged the potential use of LLMs for literature retrieval and have undertaken exploratory attempts in this area[35-37]. Yong et al. utilized ChatGPT and Microsoft Bing AI to conduct literature searches using search strategies from an existing review, with the 24 references cited in that review serving as the gold standard. The study ultimately found that of the 1,284 studies retrieved by ChatGPT, only 7 (0.5%) were directly relevant, while Bing AI retrieved 48 studies, of which 19 (40%) were directly relevant. The results suggest that the use of ChatGPT as a tool for real-time evidence generation is not yet accurate and feasible[37]. By contrast, in our study, we compared the performance of GPT-4, Claude-3, and Mistral 7×8b to generate search strategies five times, conducting independent searches with each model. As shown in **Table 3**, each search conducted by the three models achieved an accuracy rate exceeding 55%, with Claude-3 reaching as high as 90% accuracy and GPT-4 consistently achieving accuracy rates above 70%.

In the context of literature screening, previous research has conducted single-model testing (GPT-4) and multi-model comparisons (including GPT-4, GPT-3.5, Google PaLM, Meta Llama 2, FlanT5, OpenHermes-NeuralChat, Mixtral, and Platypus 2) specifically for title and abstract screening. The findings indicate that the majority of large language models (LLMs) achieved sensitivity rates exceeding 80% in abstract screening, while specificity generally remained around 30%. Notably, GPT-4 exhibited a specificity rate above 85% and demonstrated an accuracy surpassing 90%[38-42], which is consistent with our research findings. The above studies employed LLMs for title and abstract screening, demonstrating that, LLMs hold significant promise for assisting reviewer screen abstract. Building on abstract screening, our research introduced full-text screening to assess LLMs' capabilities in understanding and filtering longer texts. Qusai et al. conducted abstract and full-text screening using GPT-4 and concluded that, with highly reliable prompts, GPT-4's performance in full-text literature screening approached "human-like" levels[39]. In contrast, our study found that when conducting full-text screening, the performance of LLMs—including GPT-4, Claude-3, and Mistral 7×8b—was lower than that for abstract screening, and the differences among the three models also became less pronounced. The possible reason for this discrepancy lies in the information density and concentration in abstracts, where GPT-4 can more easily identify key information relevant to screening criteria[39]. Full-text documents contain extensive details and secondary information, making the task more complex for the model, increasing its attentional load, and making it challenging to effectively link context across different sections of the text[43].

Numerous studies have evaluated the performance of Large Language Models (LLMs) in extracting data from various types of texts, including free-text examination reports, electronic health records, and academic papers. [44-48]. Gerald, Konet, and colleagues compared the accuracy of GPT-4 and Claude-2 in extracting data from 10 articles encompassing 160 data fields. Their findings indicated that Claude-2 demonstrated high accuracy (96.3%), whereas GPT-4 with the plug-in exhibited a lower accuracy rate (68.8%)[46,47]. Khan et al. evaluated the accuracy of GPT-4-turbo and Claude3-Opus in independently extracting data from 22 articles. Each LLM generated 506 responses, and both models achieved a mean accuracy exceeding 90%[48]. However, in our study, the accuracy of GPT-4 in data extraction was only around 80%, which aligns closely with the findings reported by Khraisha[39], and the accuracy of Claude-3 was between 50% and 60%. Our findings differ substantially from those of the two previously mentioned studies. A possible reason is that in our study, the three models were accessed via Application Programming Interfaces (APIs) in independent sessions, ensuring that their data extraction accuracy was not influenced by third-party factors. In contrast, Konet used the paid browser version of GPT-4, which required a third-party plug-in to parse PDF documents. Consequently, the authors attributed most of the errors to the plug-in[47]. Additionally, the first two studies assessed LLMs by extracting only a few hundred data points, which may introduce bias in accuracy. In contrast, our study implemented a comprehensive evaluation framework requiring each model to extract 1,120 data points from 20 randomized controlled trials (RCTs), encompassing study information, intervention details, and outcome information—comprising a total of 3,360 individual data fields. To the best of our knowledge, this represents the largest-scale evaluation of LLM performance in systematic review data extraction to date.

This study has several limitations: (1) While it evaluates the performance of LLMs across the automated workflow of a systematic review (SR), only one SR type was included, necessitating further exploration across more types and studies; (2) it offers a horizontal comparison of three leading LLMs, yet does not compare them against other types of automation tools; (3) although the three models demonstrated reasonable performance, we did not employ additional techniques to enhance their capabilities across various tasks.

**Conclusion**

Large language models (LLMs) demonstrate significant potential for streamlining systematic review (SR) process. Among them, GPT-4 outperforms Claude-3 and Mistral 7×8b in overall performance and exhibits higher accuracy. Despite existing limitations such as inaccuracy, large language models hold promise as powerful assistive tools for researchers conducting SRs.

**Supplementary file 1. Quality assessment criteria for Study Design**

| Metric | Explainations |
|---|---|
| Accuracy | 1 points: Entirely erroneous and unsuitable for the research.<br><br>2 points: Mostly inaccurate, with significant errors or major misalignments with the research question.<br><br>3 points: Somewhat accurate, but contains notable inaccuracies or minor misalignments with the research question.<br><br>4 points: Mostly accurate, with only a few minor errors or slight misalignments with the research question.<br><br>5 points: Completely aligned with the research question, no errors. |
| Integrity | 1 points: Severely incomplete, lacking critical information.<br><br>2 points: Lacking some important details, with noticeable omissions that impact understanding.<br><br>3 points: Adequate in coverage, but missing some non-critical details that could provide a fuller picture.<br><br>4 points: Nearly complete, with only minor omissions or missing supplementary information.<br><br>5 points: Comprehensive and exhaustive, without omissions. |
| Relevance | 1 points: Irrelevant to the research question.<br><br>2 points: Slightly relevant, but not closely connected to the research question.<br><br>3 points: Moderately relevant, touching on the research question but not fully aligned.<br><br>4 points: Mostly relevant, with minor deviations from the research question.<br><br>5 points: Highly relevant and closely aligned with the research question. |
| Consistency | 1 points: Logically incoherent, with contradictions throughout.<br><br>2 points: Mostly inconsistent, with several contradictions or logical gaps.<br><br>3 points: Somewhat consistent, but with a few noticeable contradictions or lapses in logic.<br><br>4 points: Generally consistent, with only minor logical inconsistencies or very subtle contradictions.<br><br>5 points: Logically consistent, with no contradictions. |
| Overall performance | The average scores of the four metrics. |

**Supplementary file 2. Reliability Analysis**

**Table1. Intraclass Correlation Coefficients (ICC) for Study Design Assessment**

|  | PICO Design | Eligibility Criteria |
| --- | --- | --- |
| Accuracy | 0.857 | 0.643 |
| Integrity | 0.500 | 0.692 |
| Relevance | 0.857 | 0.750 |
| Consistency | 0.750 | 0.375 |

**Table2. Coefficient of Variation (CV) for Different Tasks Across GPT-4, Claude-3, and Mistral 7×8b Models"**

|  | GPT-4 | Claude-3 | Mistral7×8b |
| --- | --- | --- | --- |
| Search strategies | 0.073 | 0.093 | 0.136 |
| Literature screen |  |  |  |
|     Abstract screen | 0.007 | 0.003 | 0.001 |
|     Full-text screen | 0.070 | 0.000 | 0.000 |